# The quality of the Web of Science data: a longitudinal study on the completeness of authors-addresses links

Abdelghani Maddi
abdelghani.maddi@hceres.fr
Observatoire des Sciences et Techniques, Hcéres, 2 Rue Albert Einstein, Paris, 75013 France

Lesya Baudoin[1]
lesya.baudoin@inserm.fr
Observatoire des Sciences et Techniques, Hcéres, 2 Rue Albert Einstein, Paris, 75013 France

.

# Abstract

The author-affiliation links are the essential elements used for multiple purposes, such as the disambiguation of authors, the attribution of credits of a publication and fractional counting, the analysis of scientific networks, etc.

In this article we analyzed the author-affiliation link quality in the Web of Science (WoS) database between 2000 and 2021. We analyzed the link completeness for 32,676,914 scientific publications under different angles: WoS index, document type and the number of authors per publication.

The analysis showed that the author-affiliation link begins to be well informed from 2008. The share of publications for which all addresses and all authors are linked is close to 100% from 2016. The results show a strong variability according to the WoS index, the document type and the number of authors per publication. AHCI is the index with the highest completeness rate, unlike the SCI. For the document type, these are the Conference proceedings where the completeness rate is better and/or can be completed. Regarding the number of authors, statistics show that the higher the number, the more addresses and unlinked authors there are.

Finally, the analysis of a random sample of 100 publications showed that in more than 50% of the cases, the author-address links do not exist in the original publication, and the WoS reproduced only the available information provided by the editor.

# Keywords

Web of Science; metadata; authors-addresses links;data quality; bibliometrics.

# JEL codes

C8; Y1; D8

# Compliance with Ethical Standards

The authors have no relevant financial or non-financial interests to disclose.

---

[1] Present address: Institut pour la Recherche en Santé Publique BIOPARK, 8 rue de la Croix Jarry, Paris, 75013 France.

# Introduction

In the current scholarly publication practice, the authors of a research article report/declare their affiliation with the institution where the research was conducted. Institutional affiliation or institutional address is a required information for the manuscript submission in journal management systems of the most journal publishers. A common structure of a research paper includes authors' affiliations, although exceptions can be found, especially in the humanities. The paper layout is journal-specific and subject-specific; however, the relationships between authors and affiliations are generally defined by symbols including letters, numbers or special characters.

The ability to link authors to affiliations is of particular interest for bibliometric studies since it generates several relevant insights.

Author-affiliation link is an important element commonly used for author name disambiguation (Hussain and Asghar 2017; Müller et al. 2017; Reijnhoudt et al. 2014; Smalheiser and Torvik 2009; Tang and Walsh 2010). It enables to establish a reliable association between an author and an address; therefore, if two homonymous authors are related to the same affiliation, they are very likely to be the same person.

Establishing an accurate association between an author and an address is essential for studying researchers' career and mobility (Albarran et al. 2017; Deville et al. 2014; Macháček et al. 2021; Moed et al. 2013; Robinson-García et al. 2018; Sugimoto et al. 2016). This kind of analysis can reveal how often do authors change an institution or a country during their career, and identify research migration flows between countries and profiles of international mobility.

Using direct links between authors and their affiliation information opens up new avenues in the studies of research collaboration. Weighting the credits of an institutional or geographical player according to the number of affiliated authors appears fairer than simply dividing the credits regardless of how many authors of each player contribute to the research (Abramo et al. 2021; Abramo and D'Angelo 2020; CWTS 2021; Roberge et al. 2021). Linking authors and affiliations as they appear in papers allows more fine-grained analysis of a collaboration structure, such as modelling bibliometric performance of scholars from different countries or institutions (Costas and Noyons 2013), analysing how ethnic diversity in scientific collaboration affects scientific impact (Ding et al. 2021), tracking gender differences among countries in publication productivity and citation impact (Boekhout et al. 2021; Huang et al. 2020).

More generally, the possibility to assign authors to specific locations is important to address all spatial aspects of research activities, like the effects of physical distance on scientific interaction or geographical dimension of international flows of citations(Abramo et al. 2020a, 2020b; Frenken et al. 2009; Khor and Yu 2016; Tijssen et al. 2012; Waltman et al. 2011; Wuestman et al. 2019).

Besides, complete and accurate authors-addresses links are also invaluable in bibliometric analyses and research evaluation practices, and can also impact the rankings of research (Donner et al. 2020; Orduna-Malea et al. 2019).

The most of bibliographic databases provide an affiliation information, however few relate addresses to authors. Web of Science (WoS), one of the most widely used source for bibliometric studies, has introduced author-address links since 2008 for new records added to the database.[2] Completeness of the metadata is critical from the perspective of scientometric studies. Several papers have analysed the WoS data quality on several issues: accuracy of the document type assignment (Donner 2015, 2017), accuracy of citation (Buchanan 2006; Franceschini et al. 2014, 2016; van Eck and Waltman 2019), completeness of the DOI information (Gorraiz et al. 2016; Xu et al. 2019), missing author address information (Liu et al. 2018).

---

[2] https://clarivate.libguides.com/webofscienceplatform/coverage

However, to our knowledge, no analysis of completeness of the author-address links in the WoS has been published so far. In this paper, we present an empirical large-scale analysis of author-address links data quality in the WoS Core collection for the period 2000-2021. We track its evolution and point out some issues related to the errors induced in the course of data entry and processing.

# Data and method

In this paper, we used the data of the Observatoire des Sciences et Techniques (OST) in-house version of the WoS database updated in the April of 2022 and including five indexes from the WoS Core collection (table 1). The analysis includes only articles, reviews and conference proceedings documents types. The data is organized according to Oracle relational schema (queried via SQL developer). Data visualization is carried out using standard software such as Excel, XLSTAT, R and Phyton.

**Table 1: WoS indexes used in the study**

| Collection | Edition | Edition Full Name |
|---|---|---|
| WoS Core collection | SCI | Science Citation Index Expanded |
| | SSCI | Social Sciences Citation Index |
| | AHCI | Arts & Humanities Citation Index |
| | ISTP | Conference Proceedings Citation Index - Science |
| | ISSHP | Conference Proceedings Citation Index - Social Sciences |

The final dataset contains 32,676,914 papers published between 2000 and 2021. Although the last two years are incomplete, this is not problematic regarding our research question (which is the author-address link completeness analysis).

**Figure 1: distribution of papers by document type, 2000-2021**

1a) Number of publications by type | 1b) % per year

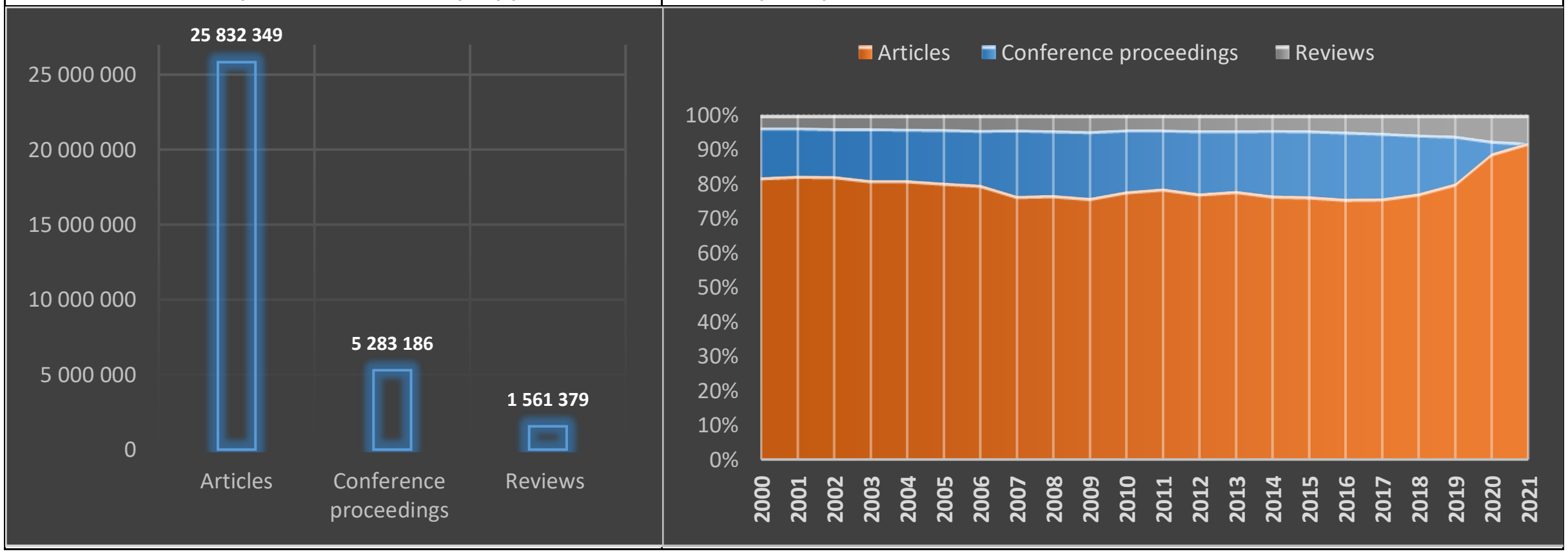

Figure 1 shows that articles represent almost 80% of the publications in the database, followed by conference proceeding (16%) and reviews (5%). We note the decrease in the share of conference proceeding over the last three years: this is an artefact of the database, since the data for this type of publication are indexed with 1-3 years delay.

To study the completeness of the author-address link, we used the data provided by *Clarivate Analytics* (https://clarivate.com/) on the author-address link on the one hand, and the bibliographic data of the publications on the other hand. Thus, for each publication we calculated the following statistics, by WoS index and by document type:

- The total number of authors;
- The total number of addresses;
- The number of linked authors;
- The number of linked addresses.

From these four statistics, we can distinguish three major situations, which are broken down into several cases summarized in Figure 2. Thus, we formed eight disjoint groups for each case.

**Figure 2: Scheme of completeness groups of author-address links**

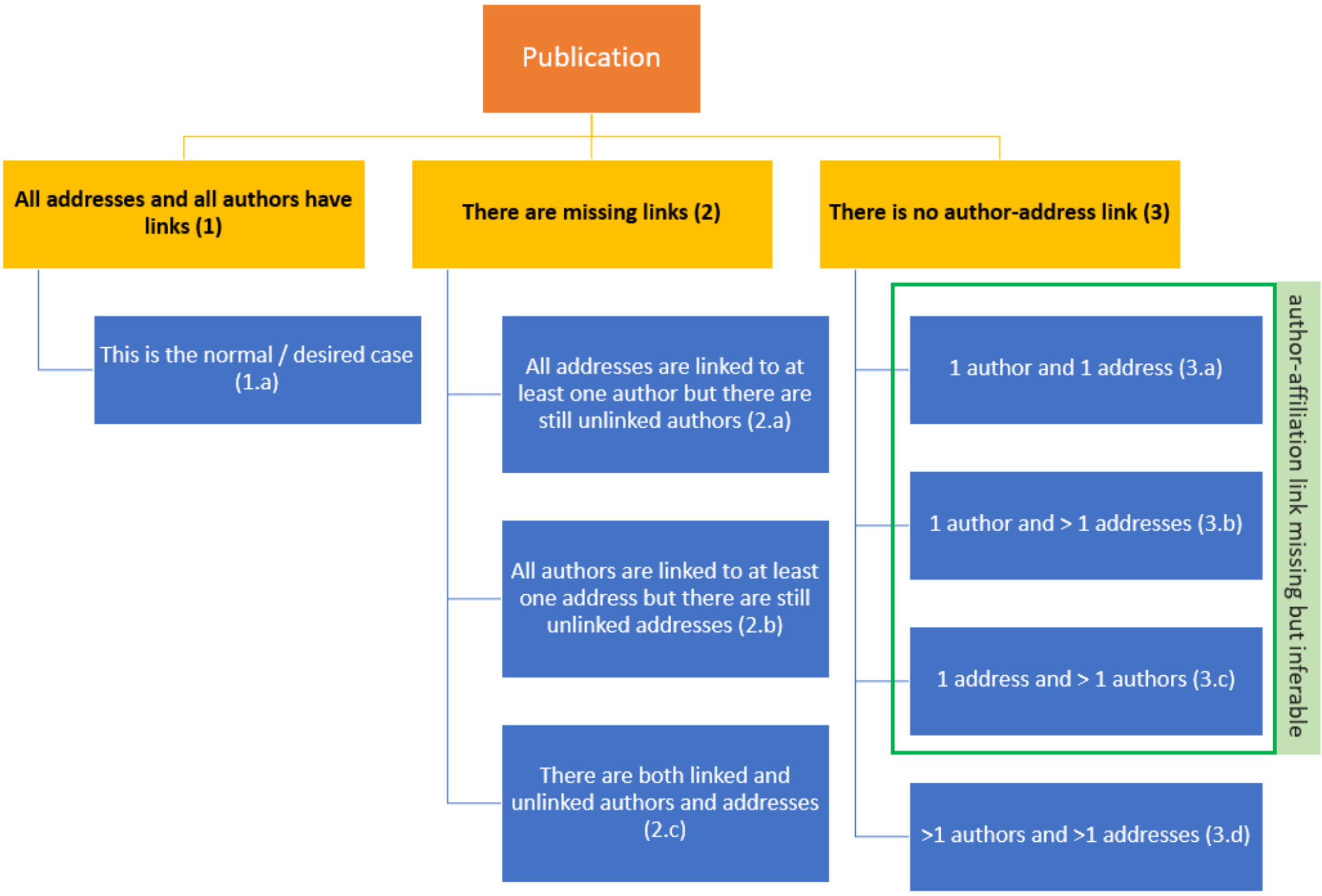

**Figure 3: An example of missing links author-address in WoS**

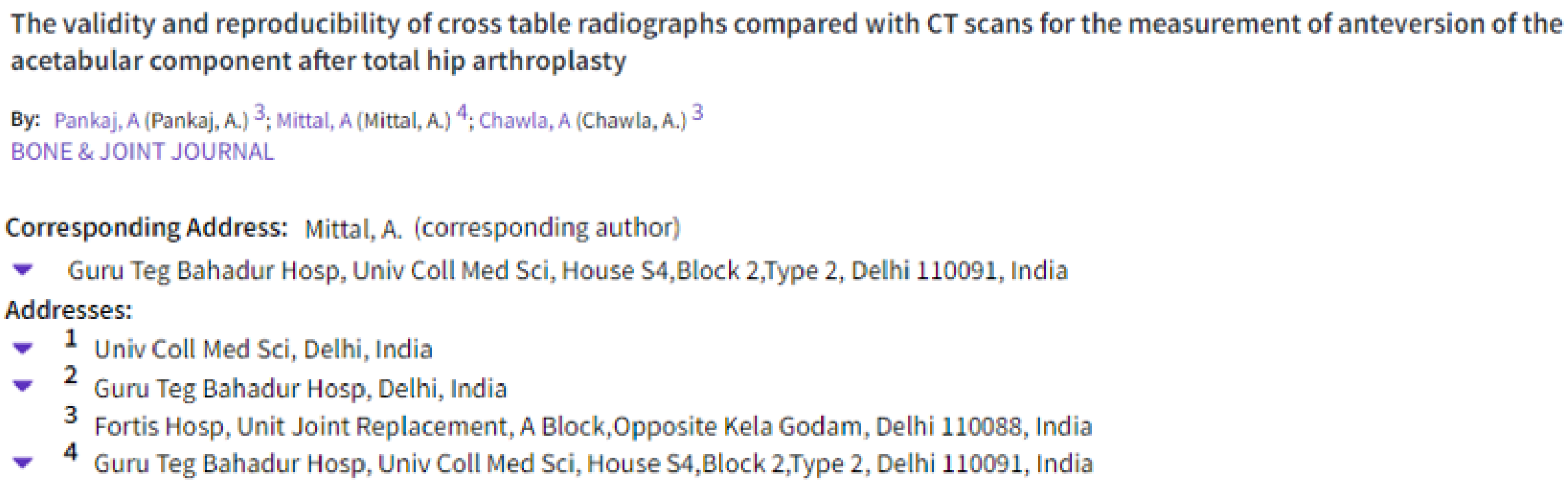

The validity and reproducibility of cross table radiographs compared with CT scans for the measurement of anteversion of the acetabular component after total hip arthroplasty

By: Pankaj, A (Pankaj, A.) [3]; Mittal, A (Mittal, A.) [4]; Chawla, A (Chawla, A.) [3]
BONE & JOINT JOURNAL

Corresponding Address: Mittal, A. (corresponding author)
Guru Teg Bahadur Hosp, Univ Coll Med Sci, House S4,Block 2,Type 2, Delhi 110091, India
Addresses:
1 Univ Coll Med Sci, Delhi, India
2 Guru Teg Bahadur Hosp, Delhi, India
3 Fortis Hosp, Unit Joint Replacement, A Block,Opposite Kela Godam, Delhi 110088, India
4 Guru Teg Bahadur Hosp, Univ Coll Med Sci, House S4,Block 2,Type 2, Delhi 110091, India

Figure 3 shows an example of missing author-address links. As we can see, addresses 1 and 2 are not linked to any author. This publication falls into the category 2.b in Figure 2.

Table 2 : **distribution of publications by type of completeness groups of author-address links**

| Code type | # | % | % |
|---|---|---|---|
| 1.a | 22 669 086 | 69,4% | 69,4% |
| 2.a | 749 646 | 2,3% | 3,2% |
| 2.b | 47 911 | 0,1% | |
| 2.c | 240 900 | 0,7% | |
| 3.a | 2 069 631 | 6,3% | 27,4% |
| 3.b | 78 146 | 0,2% | |
| 3.c | 2 943 173 | 9,0% | |
| 3.d | 3 878 421 | 11,9% | |
| **Sum** | **32 676 914** | **100,0%** | **100,0%** |

Table 2 shows the distribution of publications by type of completeness groups of author-address links (see Figure 2 for the groups description). We note that throughout the WoS database (all years, indexes and documents types), the first case (1.a) represents almost 70% of the publications. To this we can add publications with author-address link missing but inferable (cases 3a., 3.b and 3.c) which together represent about 15%. In sum, publications with links provided or that can be inferred represent 85% of WoS publications.

# Results

In this section, we present the results for each group for the WoS database as a whole, by WoS index, by document type and by the number of authors.

## The overall completeness of author-address links

Figure 4 shows, as expected, that the quality of author-address links gradually improves from 2008, year from which the WoS begins to provide this information.

**Figure 4: Evolution of the completeness of author-address links in the WoS database**

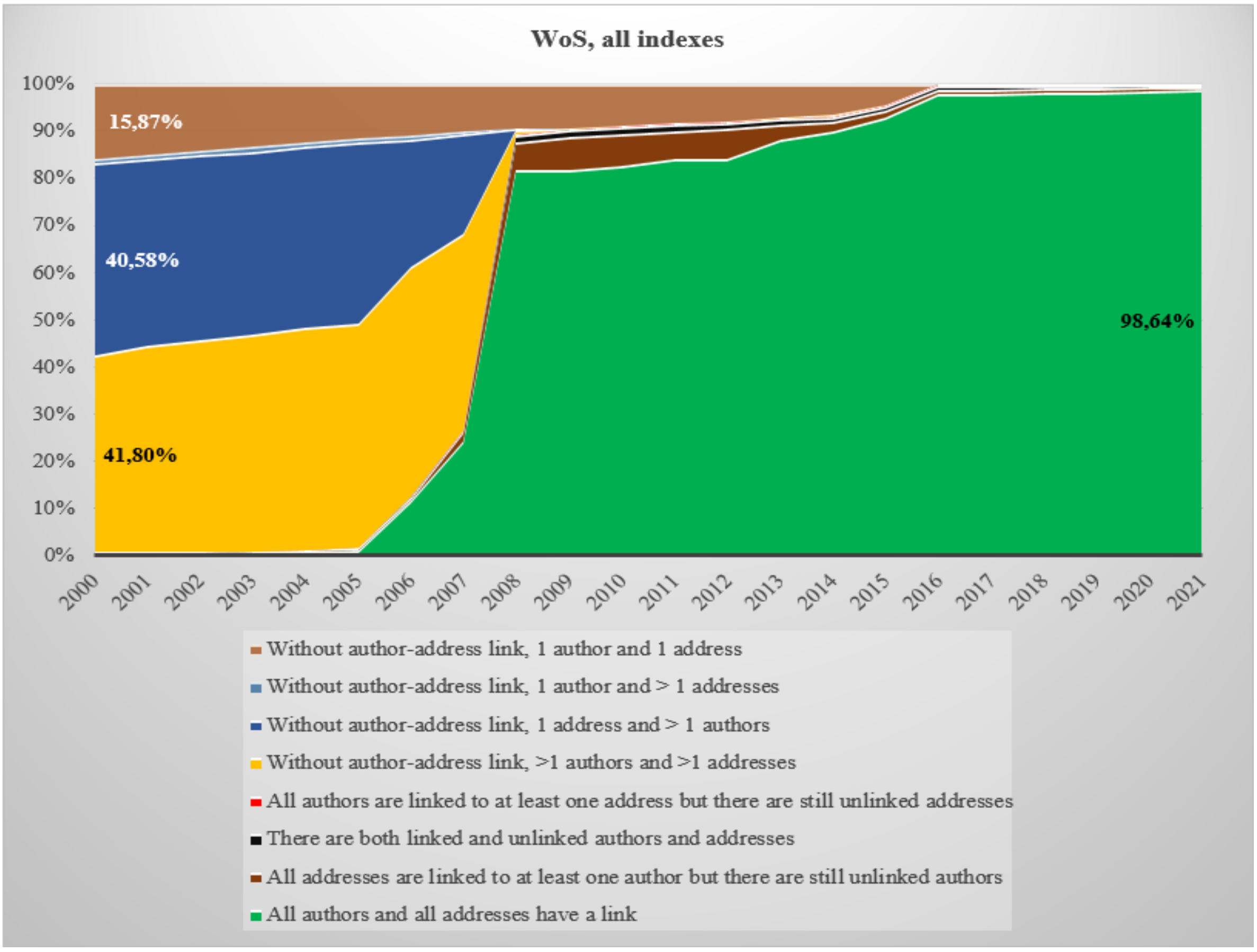

As can be seen from the data, before 2008 almost all publications have no author-address links. However, more than 55% of publications have either a single address and several authors (40.58% in 2000), or a single author and several addresses (15.87% in 2000). Consequently, for this type of publication it is quite possible to link the addresses to the authors, insofar as there cannot be any ambiguity on the author-address link. At the opposite, for publications without an author-address link which have several authors and several addresses (which represent 41.80% in 2000), it is not possible to establish a link without the information being explicitly given in the metadata. Hence, for author-address fractioning purposes, these publications would be problematic in the WoS database, insofar as only fractioning by address is possible.

Furthermore, we note that publications where all the addresses are linked to at least one author but there are still authors without a link, represent approximately 10% between 2008 and 2013, then gradually disappear.

## Completeness by WoS citation index

**Figure 5: Evolution of the completeness of author-address links in the WoS database, by edition (index)**

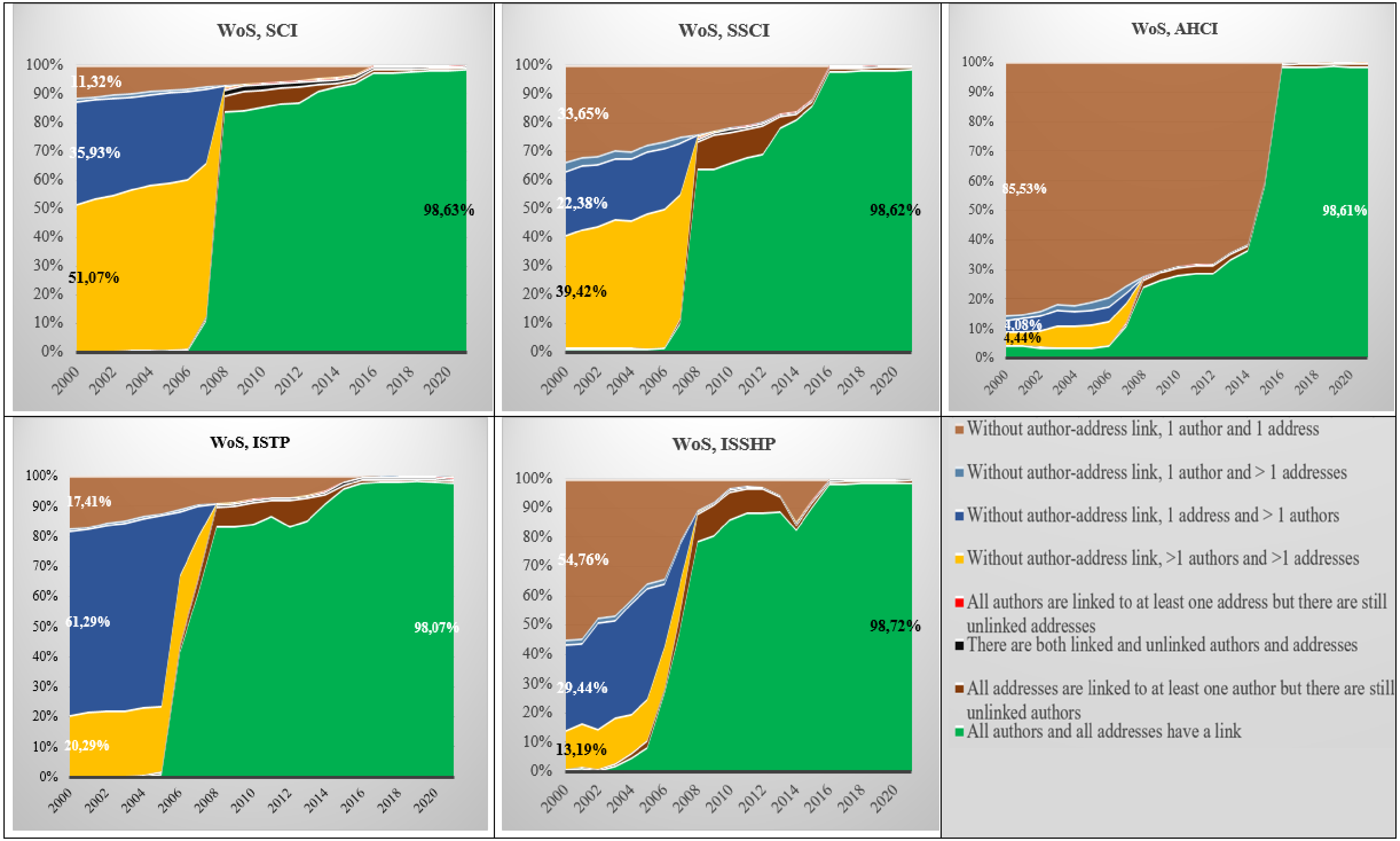

Figure 5 shows the evolution of the completeness of author-address links in the WoS database, by index. The evolution patterns are quite different depending on the editions (indexes). The share of publications without author-address link involving multiple addresses and multiple authors is higher in the SCI – Science Citation Index (about 51% in 2000). This is to be expected as this index includes publications from disciplines, such as physics or biology, with large numbers of authors and addresses. Thus, the higher is the number of authors and/or address in a publication, the higher is the probability of having missing links. In contrast, the share of publications without author-address link, involving a single author and a single address is relatively low in SCI, with only about 11% in 2000. This is the lowest share for this group of publications, among all WoS indexes.

In the Social Sciences Citation Index - SSCI, the proportions are quite different from those observed for SCI, especially regarding publications without author-address link with a single author and a single address (which share is three times higher than that of the SCI in 2000). This is explained by the disciplinary orientation of this index, which includes the social sciences with a significant share of publications without collaboration. This specificity is much more marked in the Arts & Humanities Citation Index - AHCI where the majority of publications (over 85% in 2000) are without author-address link, with a single author and a single address. It can be seen in Figure 5 that the link remains missing for these publications until 2015. Note that on this index (AHCI) it is possible to infer author-address links for the majority of publications, even before 2008, given the high share of publications with a single address and/or a single author. The share of publications with missing author-address links is only 4.44% in 2000. The share of publications where all addresses are linked to at least one author but there are still unliked authors, is also very low for AHCI (compared to SCI and SSCI).

With regard to the two indexes of proceedings (ISTP - Conference Proceedings Citation Index - Science, and ISSHP - Conference Proceedings Citation Index - Social Sciences), there are two different structures. As expected, the evolution observed for the ISSHP index is similar to that of SSCI and AHCI. Since ISSHP mainly indexes conferences proceedings in the social sciences and humanities, the share of (unlinked) publications with a single author and/or a single address is relatively high. This type of publications represents almost 55% in 2000. Furthermore, it should be noted that in 2014, the ISSHP index recorded a peak in publications with a single author and a single address (without link). Further analysis should be done to be able to explain it.

For ISTP, publications (without link) with an address and more than one author dominate with a share of over 61%, the highest share for this type of publications in comparison with all the other indexes. Similarly, the share of publications without author-address link involving several authors and several addresses is higher than that of the ISSHP (20.29% against 13.19%).

In short, we note that the SCI index is the one that contains the most publications without an author-address link that cannot be "corrected", i.e. for which the author-address link cannot be inferred because of the multiplicity of authors and addresses per publication (this group of publications represents 51% in 2000). This is also the case for SSCI, but to a lesser extent (this group of publications represents 39% in 2000 in this index). On the contrary, due to the nature of the humanities and social sciences, characterized by a relatively high share of single author and single address publications (especially in the early 2000s), author-address links, even if missing or non-existent, can be identified and completed for the majority of publications before 2008.

Completeness by document type

**Figure 6: Evolution of the completeness of author-address links in the WoS database, by document type**

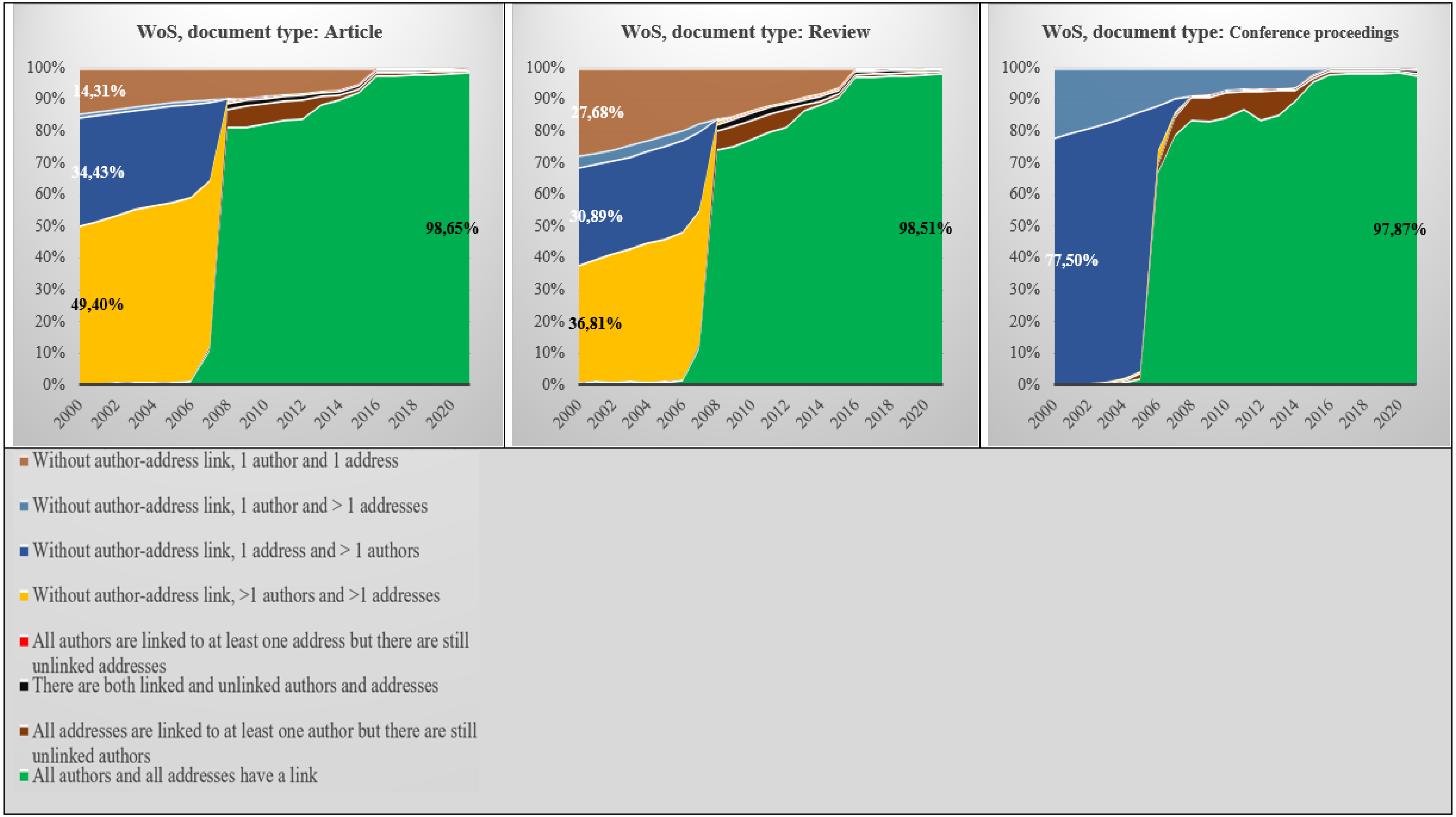

Figure 6 shows the evolution of the completeness of author-address links in the WoS database by document type. As can be seen, the distribution of publications according to the completeness of author-address links is similar to that observed for the entire database, especially for the "article" document type. This is obviously to be expected insofar as this type of document (Article) represents nearly 80-90% of publications. The evolution of author-address completeness for the document type "Review" is quite interesting. It shows that publications with a single author and a single address represent almost a third of publications at the start of the period. These are overall higher than for the other two document types over the period. This result is quite logical for this document type (Review) which does not necessarily require collaboration between researchers or institutions.

For the "Conference proceedings" document type, the distribution of publications is quite different from that of "Articles" and "Reviews". The share of publications with a single address and several authors (or a single author and several addresses) is close to 100% between 2000 and 2006. The explanation for this distribution can be found in the fact that, in general, in the scientific conferences, the institutional dimension is prevalent. Thus, it is common in conferences that one or more researchers come to represent their institutions and the progress of their research. Thus, conferences are a place that can lead to future collaborations between researchers from several institutions working on the same topics. Since the majority of publications for this document type are publications without an author-address link, involving a single address and several authors (or a single author and several addresses), it is possible to improve the quality of the data by linking authors and addresses as indicated above.

Figure 7 shows the evolution of the completeness of author-address links in the WoS database by authors number. We note that for publications with a single author and a single address, the link is not entered in the database for the majority of publications, and this until 2016. This result is quite curious insofar as this information on the author-address link has been implemented since 2008. However, as stated above, this is not a problematic for this type of publications because the link can be established without ambiguity. The same goes for publications involving a single address and several authors which represent 55.18% in 2000 for the category 2 to 5 authors, 25.91% for the category 6 to 10 authors and 19.95% for the category +10 authors. Otherwise, we also note that publications without an author-address link involving a single author and several addresses have a fairly low share between 2000 and 2008 and are gradually disappearing.

Furthermore, Figure 7 shows that the more the number of authors per publication increases, the more the quality of the author-address links deteriorates. Which is to be expected, as the likelihood of having missing links increases with the number of authors and addresses. Thus, publications without an author-address link involving several authors and several addresses represent 44.42% in 2000 for publications involving 2 to 5 authors, 73.81% for publications involving 6 to 10 authors, and 80.08% for publications involving more than 10 authors. The missing links for these publications cannot be corrected (or completed) afterwards. The same applies to publications for which all the addresses are linked to at least one author but there are still authors without an author-address link. These publications appeared after 2008 and their share is gradually declining. In 2008, they represent about 5% for the 2 to 5 authors category, about 10% for the 6 to 10 authors category and nearly 20% for the +10 authors category.

Completeness according to the authors number

**Figure 7: Evolution of the completeness of author-address links in the WoS database, by authors number**

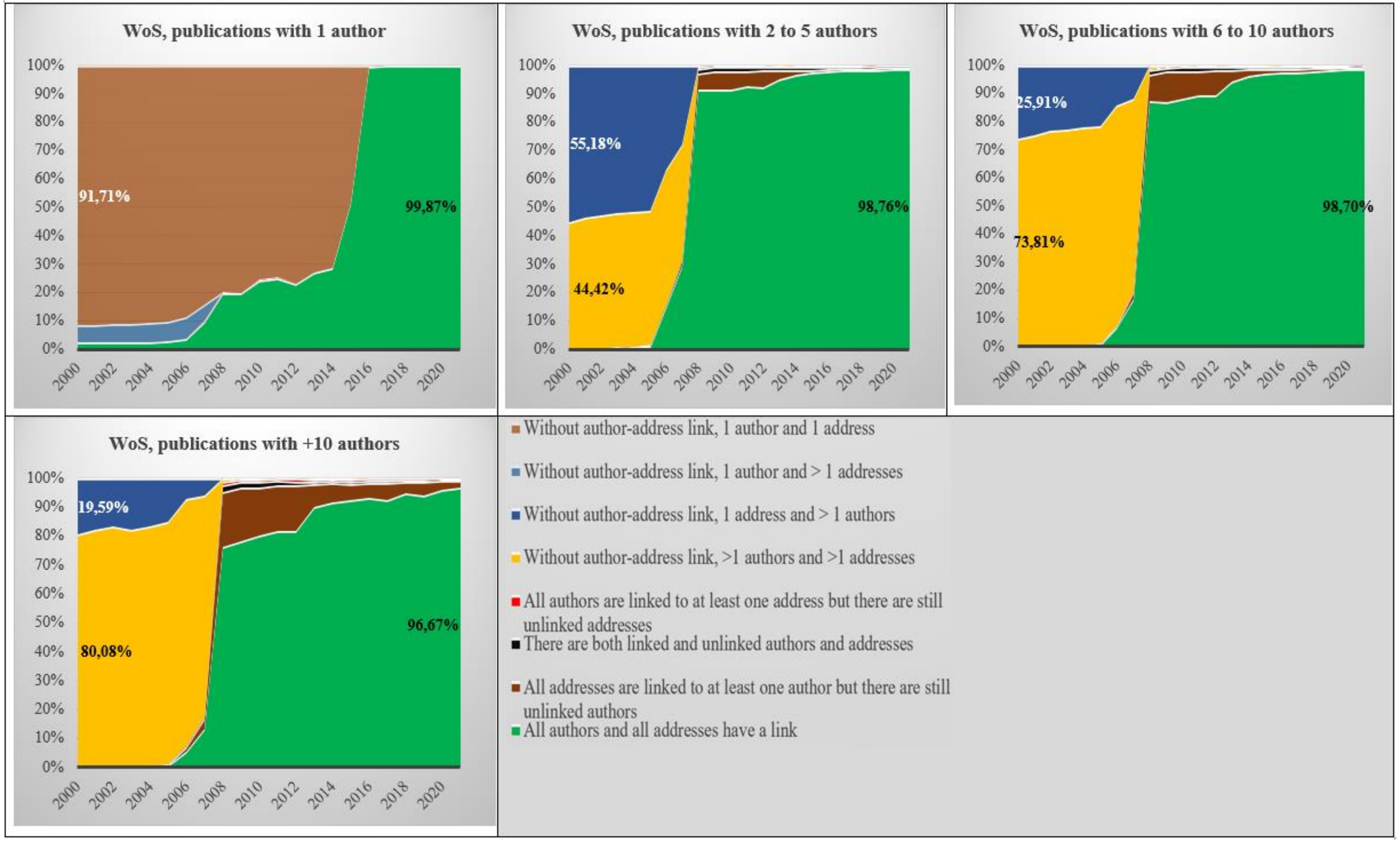

## Publications with complete author-address link or with an author or an address

Figure 8 shows the evolution of the completeness of author-address links by index and for the whole WoS database, including publications without link involving a single author or a single address. It can be seen that the evolution of the Social Sciences Citation Index - SSCI is very close to that of the WoS. Thus, the data quality for SSCI and for the WoS in general is modest before 2008, then gradually improves to reach 90% complete author-address links from 2013. The rate approaches 100% from 2016.

The index with the most incomplete author-address data is the Science Citation Index – SCI with a declining rate between 2000 and 2006 (due to increased collaboration in science). Since SCI is the index with the most publications, it mechanically pulls the WoS database average down for data before 2008.

Arts & Humanities Citation Index – AHCI is the best in terms of completeness of author-address links. This result is due to the characteristics of the disciplines included in the latter (Humanities) where the number of authors and/or addresses per publication is relatively low. Thus, the rate of author-address links that are complete (or can be easly completed) exceeds 90% over the entire period to reach nearly 100% in 2021.

The two indexes of Conferences proceedings (ISTP: Conference Proceedings Citation Index – Science, and ISSHP: Conference Proceedings Citation Index – Social Sciences) have similar characteristics to AHCI in terms of the number of authors per publication. Consequently, the completeness rates of author-address links for the latter are also quite high over the entire period. Without surprise, ISSHP has a higher completeness rate than ISTP.

**Figure 8: Publications with complete author-address link or with** author-affiliation link missing but inferrable

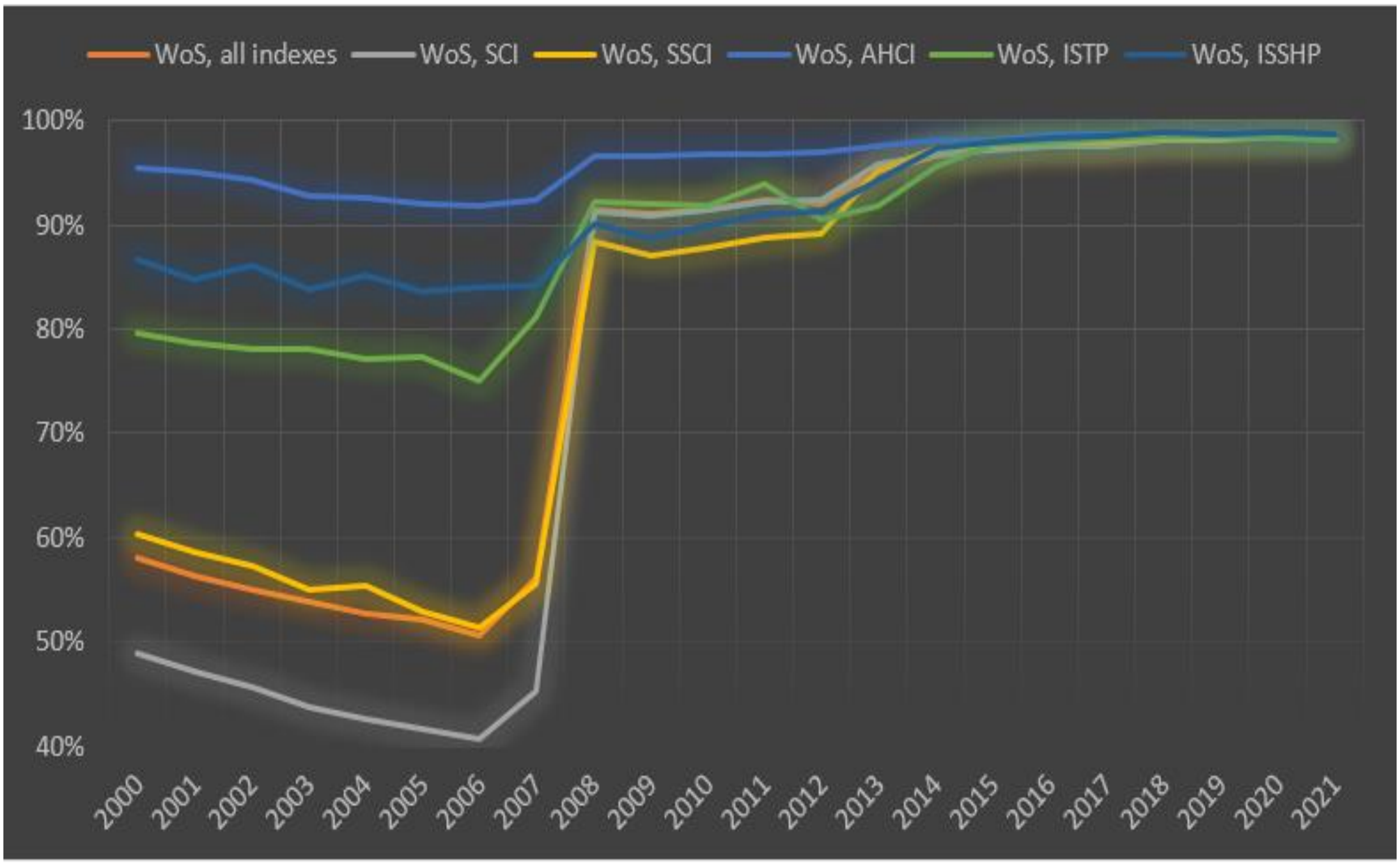

Although we use the term "completeness", it refers to the mere presence or absence of links in the WoS data. However, the causes of the incompleteness of author-affiliation links can vary: in some cases the links are not provided in the original papers, in other cases they exist but are badly captured in WoS.

To illustrate this, we drew a random sample of 100 publications. For each publication, we checked on the publisher's site and/or in the PDF of the publication whether the author-address links exist. The results show that in more than 50% of the cases, the link does not exist in the original publication and the WoS has only reproduced the available information provided by the publisher.

Furthermore, it should be noted that in the other half where the author-address link exists, we have found that in the majority of cases the link is established with special characters (asterisks, crosses, etc.), or else below each author's name. One wonders if an adjustment of the algorithm of identification of the links author-address of WoS would make it possible to recover this type of missing links.

## Conclusion and discussion

We analysed the completeness of author-address links in the Web of Science database between 2000 and 2020, by index and by document type. As expected, papers with author-address links appear massively (>80%) in the database from 2008, the year in which an editorial decision was taken by Clarivate Analytics. In previous publication years, a sizable proportion of papers with author-address links can be observed, in decreasing order: 24% in 2007, 12% in 2006, 1% in 2005, then <1% in earlier years. Presumably, these papers were added to the database in 2008 or later. The proportion of papers with author-address links in the database is growing slowly from year to year; it reaches 90% in 2014. The completeness can be considered as especially good from 2016 (98%).

Some combinations of authors number - addresses number make it possible to identify non-ambiguous associations between authors and affiliations, even without explicit links. We distinguish to this end 3 following classes of papers without author-address links: single-author single-address publications (1-1), single-author and several addresses (1-∞), and several authors and single address (∞-1). It is assumed that the author-affiliation links can be inferred from the relationships existing in these classes. When taking together the papers of these tree classes with those having compete author-address links, the associations between authors and affiliations can be identified for more than half publications before 2008, and more than 90% beginning from 2008. However, some reservations may be expressed regarding the single-address papers with dozens of authors. In most cases they result from the way Clarivate Analytics processes the "group authors"[3]. According to the WoS indexing policy, the individual members of a study group are added in the byline of the Author field, in the same way as the effective authors. Since the affiliation information for the group members is often sketchy, if exists, this information is not captured, and therefore only the corresponding address is provided. Furthermore, adding the group members into the authors byline often the source of other anomalies such as duplicate authors or missing affiliation links. The soundness of this indexing policy can be questioned, at least in regard of biomedical publications where the role of authors is guided by principles set out by ICMJE[4].

[3] https://support.clarivate.com/ScientificandAcademicResearch/s/article/Web-of-Science-Core-Collection-Group-Author-field-definition-and-indexing-policy?language=en_US

Analysis of the publication classes defined according to the number of authors vs. the number of affiliations is echoed in a certain number of studies addressing the evolution of authorship patterns in scholarly literature. The dramatic decrease of the single-authored papers in recent years is simultaneous with the growing trend for teamwork leading to increase in a number of authors per paper (Abt 2007, 2017; Adams et al. 2019; Chinchilla-Rodriguez et al. 2021; Greene 2007; Larivière et al. 2015; Nabout et al. 2015; Weeks et al. 2004). The proportion of single-authored papers in our data is similar to those found in WoS by Larivière et al. (2015) : from around 15% at the beginning of 2000 and decreasing onward. However it differs significantly from the proportion reported by Chinchilla-Rodriguez et al. (2021) which accounts for as high as 46%. The proportion of single-address papers observed in our study declining steadily from 40% in 2000, is in line with Larivière et al. (2015).

The proportion of paper with incomplete links (all types of incompleteness taken together) remains quite high (8% in average) in the first 5 years (2008-2012) following the introducing the author-address links in the database, then gradually decreases until it reaches below 1.5% in 2020. This evolution can possible be explained by improvements in data capture and processing. The remaining 1.5% of papers with incomplete links may be due both to the databases errors and to the insufficient quality of the data provided by publishers.

# Bibliography

Abramo, G., & D’Angelo, C. A. (2020). The domestic localization of knowledge flows as evidenced by publication citation: the case of Italy. *Scientometrics*, *125*(2), 1305–1329. https://doi.org/10.1007/s11192-020-03487-5

Abramo, G., D’Angelo, C. A., & Di Costa, F. (2020a). Knowledge spillovers: Does the geographic proximity effect decay over time? A discipline-level analysis, accounting for cognitive proximity, with and without self-citations. *Journal of Informetrics*, *14*(4), 101072. https://doi.org/10.1016/j.joi.2020.101072

Abramo, G., D’Angelo, C. A., & Di Costa, F. (2020b). The role of geographical proximity in knowledge diffusion, measured by citations to scientific literature. *Journal of Informetrics*, *14*(1), 101010. https://doi.org/10.1016/j.joi.2020.101010

Abramo, G., D’Angelo, C. A., & Di Costa, F. (2021). On the relation between the degree of internationalization of cited and citing publications: A field level analysis, including and

[4] http://www.icmje.org/recommendations/browse/roles-and-responsibilities/defining-the-role-of-authors-and-contributors.html

excluding self-citations. *Journal of Informetrics*, *15*(1), 101101. https://doi.org/10.1016/j.joi.2020.101101

Abt, H. A. (2007). The future of single-authored papers. *Scientometrics*, *73*(3), 353–358. https://doi.org/10.1007/s11192-007-1822-9

Abt, H. A. (2017). Citations and author numbers in six sciences. *Scientometrics*, *111*(3), 1861–1867. https://doi.org/10.1007/s11192-017-2360-8

Adams, J., Pendlebury, D., Potter, R., & Szomszor, M. (2019). *Multi-authorship and research analytics* (No. 6) (p. 20). Institute for Scientific information. https://clarivate.com/webofsciencegroup/wp-content/uploads/sites/2/dlm_uploads/2019/12/WS419558643_ISI_Global_Research_Report_6_v9_RGB_SP.pdf

Albarran, P., Carrasco, R., & Ruiz-Castillo, J. (2017). Geographic mobility and research productivity in a selection of top world economics departments. *Scientometrics*, *111*(1), 241–265. https://doi.org/10.1007/s11192-017-2245-x

Boekhout, H., van der Weijden, I., & Waltman, L. (2021). Gender differences in scientific careers: A large-scale bibliometric analysis. In W. Glanzel, S. Heeffer, P. S. Chi, & R. Rousseau (Eds.), *18th International Conference on Scientometrics & Informetrics (issi2021)* (pp. 145–156). Leuven: Int Soc Scientometrics & Informetrics-Issi. https://www.webofscience.com/wos/woscc/full-record/WOS:000709638700017. Accessed 10 January 2022

Buchanan, R. A. (2006). Accuracy of cited references: The role of citation databases. *College & Research Libraries*, *67*(4), 292–303.

Chinchilla-Rodriguez, Z., Liu, J., & Bu, Y. (2021). Patterns of Knowledge Diffusion via Research Collaboration on a Global Level. In W. Glanzel, S. Heeffer, P. S. Chi, & R. Rousseau (Eds.), *18th International Conference on Scientometrics & Informetrics (issi2021)* (pp. 269–280). Leuven:

Int Soc Scientometrics & Informetrics-Issi. https://www.webofscience.com/wos/woscc/full-record/WOS:000709638700030. Accessed 17 January 2022

Costas, R., & Noyons, E. (2013). *Detection of different types of 'talented' researchers in the Life Sciences through bibliometric indicators: methodological outline* (No. CWTS-WP-2013-006) (p. 25). Leiden, The Netherlands: Centre for Science and Technology Studies (CWTS). https://scholarlypublications.universiteitleiden.nl/access/item%3A2885710/view

CWTS, C. for S. and T. (2021). CWTS Leiden Ranking. *CWTS Leiden Ranking*. Centre for Science and Technology Studies (CWTS). http://www.leidenranking.com. Accessed 10 January 2022

Deville, P., Wang, D., Sinatra, R., Song, C., Blondel, V. D., & Barabási, A.-L. (2014). Career on the Move: Geography, Stratification and Scientific Impact. *Scientific Reports*, *4*(1), 4770. https://doi.org/10.1038/srep04770

Ding, J., Shen, Z., Ahlgren, P., Jeppsson, T., Minguillo, D., & Lyhagen, J. (2021). The link between ethnic diversity and scientific impact: the mediating effect of novelty and audience diversity. *Scientometrics*, *126*(9), 7759–7810. https://doi.org/10.1007/s11192-021-04071-1

Donner, P. (2015). Document type assignment accuracy in citation index data sources. In A. A. Salah, Y. Tonta, A. a. A. Salah, C. Sugimoto, & U. Al (Eds.), *Proceedings of Issi 2015 Istanbul: 15th International Society of Scientometrics and Informetrics Conference* (pp. 1271–1272). Leuven: Int Soc Scientometrics & Informetrics-Issi.

Donner, P. (2017). Document type assignment accuracy in the journal citation index data of Web of Science. *Scientometrics*, *113*(1), 219–236. https://doi.org/10.1007/s11192-017-2483-y

Donner, P., Rimmert, C., & van Eck, N. J. (2020). Comparing institutional-level bibliometric research performance indicator values based on different affiliation disambiguation systems. *Quantitative Science Studies*, *1*(1), 150–170. https://doi.org/10.1162/qss_a_00013

Franceschini, F., Maisano, D., & Mastrogiacomo, L. (2014). Scientific journal publishers and omitted citations in bibliometric databases: Any relationship? *Journal of Informetrics*, *8*(3), 751–765. https://doi.org/10.1016/j.joi.2014.07.003

Franceschini, F., Maisano, D., & Mastrogiacomo, L. (2016). Empirical analysis and classification of database errors in Scopus and Web of Science. *Journal of Informetrics*, *10*(4), 933–953. https://doi.org/10.1016/j.joi.2016.07.003

Frenken, K., Hardeman, S., & Hoekman, J. (2009). Spatial scientometrics: Towards a cumulative research program. *Journal of Informetrics*, *3*(3), 222–232. https://doi.org/10.1016/j.joi.2009.03.005

Gorraiz, J., Melero-Fuentes, D., Gumpenberger, C., & Valderrama-Zurián, J.-C. (2016). Availability of digital object identifiers (DOIs) in Web of Science and Scopus. *Journal of Informetrics*, *10*(1), 98–109. https://doi.org/10.1016/j.joi.2015.11.008

Greene, M. (2007). The demise of the lone author. *Nature*, *450*(7173), 1165–1165. https://doi.org/10.1038/4501165a

Huang, J., Gates, A. J., Sinatra, R., & Barabasi, A.-L. (2020). Historical comparison of gender inequality in scientific careers across countries and disciplines. *Proceedings of the National Academy of Sciences of the United States of America*, *117*(9), 4609–4616. https://doi.org/10.1073/pnas.1914221117

Hussain, I., & Asghar, S. (2017). A survey of author name disambiguation techniques: 2010-2016. *Knowledge Engineering Review*, *32*, e22. https://doi.org/10.1017/S0269888917000182

Khor, K. A., & Yu, L.-G. (2016). Influence of international co-authorship on the research citation impact of young universities. *Scientometrics*, *107*(3), 1095–1110. https://doi.org/10.1007/s11192-016-1905-6

Larivière, V., Gingras, Y., Sugimoto, C. R., & Tsou, A. (2015). Team size matters: Collaboration and scientific impact since 1900: On the Relationship Between Collaboration and Scientific Impact Since 1900. *Journal of the Association for Information Science and Technology*, *66*(7), 1323–1332. https://doi.org/10.1002/asi.23266

Liu, W., Hu, G., & Tang, L. (2018). Missing author address information in Web of Science—An explorative study. *Journal of Informetrics*, *12*(3), 985–997. https://doi.org/10.1016/j.joi.2018.07.008

Macháček, V., Srholec, M., Ferreira, M. R., Robinson-Garcia, N., & Costas, R. (2021). Researchers' institutional mobility: bibliometric evidence on academic inbreeding and internationalization. *Science and Public Policy*.

Moed, H. F., Aisati, M., & Plume, A. (2013). Studying scientific migration in Scopus. *Scientometrics*, *94*(3), 929–942. https://doi.org/10.1007/s11192-012-0783-9

Müller, M.-C., Reitz, F., & Roy, N. (2017). Data sets for author name disambiguation: an empirical analysis and a new resource. *Scientometrics*, *111*(3), 1467–1500. https://doi.org/10.1007/s11192-017-2363-5

Nabout, J. C., Parreira, M. R., Teresa, F. B., Carneiro, F. M., da Cunha, H. F., Ondei, L. de S., et al. (2015). Publish (in a group) or perish (alone): the trend from single- to multi-authorship in biological papers. *Scientometrics*, *102*(1), 357–364. https://doi.org/10.1007/s11192-014-1385-5

Orduna-Malea, E., Aytac, S., & Tran, C. Y. (2019). Universities through the eyes of bibliographic databases: a retroactive growth comparison of Google Scholar, Scopus and Web of Science. *Scientometrics*, *121*(1), 433–450. https://doi.org/10.1007/s11192-019-03208-7

Reijnhoudt, L., Costas, R., Noyons, E., Börner, K., & Scharnhorst, A. (2014). 'Seed + expand': a general methodology for detecting publication oeuvres of individual researchers. *Scientometrics*, *101*(2), 1403–1417. https://doi.org/10.1007/s11192-014-1256-0

Roberge, G., Bédard-Vallée, A., & Rivest, M. (2021). *Bibliometrics Indicators for the Science and Engineering Indicators 2022: Technical Documentation* (p. 43). Science-Metrix. https://www.science-metrix.com/wp-content/uploads/2021/10/Technical_Documentation_Bibliometrics_SEI_2022_2021-09-14.pdf

Robinson-García, N., Sugimoto, C. R., Murray, D., Yegros-Yegros, A., Larivière, V., & Costas, R. (2018). Scientific mobility indicators in practice: International mobility profiles at the country level. *arXiv preprint arXiv:1806.07815*.

Smalheiser, N. R., & Torvik, V. I. (2009). Author name disambiguation. *Annual Review of Information Science and Technology*, *43*(1), 1–43. https://doi.org/10.1002/aris.2009.1440430113

Sugimoto, C. R., Robinson-Garcia, N., & Costas, R. (2016). Towards a global scientific brain: Indicators of researcher mobility using co-affiliation data. *arXiv:1609.06499 [cs]*. http://arxiv.org/abs/1609.06499. Accessed 22 December 2021

Tang, L., & Walsh, J. P. (2010). Bibliometric fingerprints: name disambiguation based on approximate structure equivalence of cognitive maps. *Scientometrics*, *84*(3), 763–784. https://doi.org/10.1007/s11192-010-0196-6

Tijssen, R. J., Waltman, L., & van Eck, N. J. (2012). Research collaboration and the expanding science grid: Measuring globalization processes worldwide. *arXiv preprint arXiv:1203.4194*.

van Eck, N. J., & Waltman, L. (2019). Accuracy of citation data in Web of Science and Scopus. *arXiv:1906.07011 [cs]*. http://arxiv.org/abs/1906.07011. Accessed 19 June 2019

Waltman, L., Tijssen, R. J., & van Eck, N. J. (2011). Globalisation of science in kilometres. *Journal of informetrics*, *5*(4), 574–582.

Weeks, W. B., Wallace, A. E., & Kimberly, B. C. S. (2004). Changes in authorship patterns in prestigious US medical journals. *Social Science & Medicine*, *59*(9), 1949–1954. https://doi.org/10.1016/j.socscimed.2004.02.029

Wuestman, M. L., Hoekman, J., & Frenken, K. (2019). The geography of scientific citations. *Research Policy*, *48*(7), 1771–1780. https://doi.org/10.1016/j.respol.2019.04.004

Xu, S., Hao, L., An, X., Zhai, D., & Pang, H. (2019). Types of DOI errors of cited references in Web of Science with a cleaning method. *Scientometrics*, *120*(3), 1427–1437. https://doi.org/10.1007/s11192-019-03162-4